\documentclass[12pt]{article}
\usepackage{amssymb}

\renewcommand{\t}{\theta}

\def\be{\begin{equation}}
\def\ee{\end{equation}}
\def\ba{\begin{eqnarray}}
\def\ea{\end{eqnarray}}
\newcommand{\nn}{\nonumber}
\newcommand{\no}{\nonumber \\}
\def\al {\alpha}
\def\lam {\Lambda}
\def\dela{\delta_{\al}}
\def\delb{\delta_{\beta}}
\def\delv{\delta_{v}}
\def\dell {\delta_{\lam}}
\def\lama {\lam_{\al}}
\def\lamb{\lam_{\beta}}

\def\della{\delta_{\lama}}
\def\dellb{\delta_{\lamb}}
\def\starcom {\stackrel{\star}{,}}
\def\L {\Lambda}

\renewcommand{\k}{\ell}
\renewcommand{\d}{\delta}
\def\D {\Delta}
\def\beq{\begin{equation}}
\def\eeq{\end{equation}}
\def\del{\partial}
\newcommand{\p}{^\prime}

\def\on{^{(n)}}

\begin{document}
\begin{titlepage}
May 2001         \hfill
\vskip -0.55cm 
\hfill hep-th/0105192

\hfill    UCB-PTH-01/20  

\hfill  LBNL-48161  
\begin{center}

\vskip .15in

\renewcommand{\thefootnote}{\fnsymbol{footnote}}
{\large \bf A Cohomological Approach to the Non-Abelian Seiberg-Witten Map}
\vskip .25in
D. Brace\footnote{email address: dmbrace@lbl.gov},
B. L. Cerchiai\footnote{email address: BLCerchiai@lbl.gov}, 
A. F. Pasqua\footnote{email address: pasqua@socrates.berkeley.edu},
U. Varadarajan\footnote{email address: udayv@socrates.berkeley.edu},
B. Zumino\footnote{email address: zumino@thsrv.lbl.gov}
\vskip .25in

{\em    Department of Physics  \\
        University of California   \\
                                and     \\
        Theoretical Physics Group   \\
        Lawrence Berkeley National Laboratory  \\
        University of California   \\
        Berkeley, California 94720}
\end{center}
\vskip .25in

\begin{abstract}

We present a cohomological method for obtaining the non-Abelian
Seiberg-Witten map for any gauge group and 
to any order in $\theta$. By introducing a ghost field, we are able to express the equations
defining the Seiberg-Witten map
through a coboundary operator, so that they can be solved by constructing a
corresponding homotopy operator.

\end{abstract}
\end{titlepage}

\newpage

\section{Introduction}
\setcounter{footnote}{0}

In the last few years noncommutative geometry has found physical
realizations in string theory as argued originally in \cite{CDS}.
Based on the existence of different regularization procedures in
string theory, Seiberg and Witten \cite{SW} claimed that
certain noncommutative gauge theories are equivalent to
commutative ones. In particular, they argued that there exists a
map from a commutative gauge field to a noncommutative one, which
is compatible with the gauge structure of each. 
This map has become known as the
Seiberg-Witten (SW) map. In this paper, we give a method for explicitly finding this map.
We will consider gauge theories on the noncommutative space defined by
\be
\left [x^i \starcom x^j \right ] =i\theta ^{ij}~,
\ee 
where $\theta$ is a constant Poisson tensor. Then the  ``$\star$'' 
operation is the associative Weyl-Moyal product
\be
\label{WMproduct}
f \star g= f ^{\frac{i}{2}\theta ^{ij}\stackrel{\leftarrow}{\partial _i}\stackrel{\rightarrow}{\partial_j}}g~.
\ee
We believe that our methods are much more general, and can in fact be
used even when $\theta$ is not constant.

In the next section, we review the methods developed in~\cite{Wess}, which
provide an essential starting point for our work.
In Section $3$ we replace the gauge parameters appearing in the SW map
with a ghost field, which makes explicit a cohomological structure
underlying the SW map. In Section $4$
we define a homotopy operator, which can be used
to explicitly write down the SW map order by order in $\theta$.
In Section $5$, we discuss some complications that arise in this
formalism and some methods to overcome them. Finally, in Section $6$
we apply our methods to calculate some low order terms of the
SW map. An Appendix contains some expansions of the star product, that
will be useful in the rest of the paper.

\section{General Review}
\setcounter{equation}{0}

In this section, we review the formalism developed in~\cite{Wess}, which
provides an alternative method for obtaining an expression for the SW map.


The original equation which defines the SW map \cite{SW} arises from the
requirement that gauge invariance be preserved in the following sense.
Let $a_i$, $\al$ be the gauge field and gauge parameter 
of the commutative theory and similarly let $A_i$, $\lam$ be the gauge 
field and gauge parameter of the noncommutative theory. 
Under an infinitesimal gauge transformation,
\beq
\label{agauge}
\dela a_i=\partial _i \al -i[a_i,\al],
\eeq
\beq
\label{eqfora}
\dell A_i=\partial_i \lam -i[A_i \starcom \lam] \equiv \partial_i \lam -i
\left ( A_i \star \lam -\lam \star A_i \right).
\eeq
Then, the SW map is found by requiring
\beq
\label{SW}
A_i+ \dell A_i=A_i(a_j+\dela a_j, \cdots).
\eeq
In order to satisfy~(\ref{SW}) the noncommutative gauge field and
gauge parameter must have the following functional dependence.
\beq
\label{map1}
\begin{array}{ll}
A_i=A_i(\,a,\partial a, \partial^2 a, \cdots)\\
\lam=\lam (\,\al,\partial \al, \cdots, a, \partial a, \cdots),
\end{array}
\eeq
where the dots indicate higher derivatives. It must be emphasized that
a SW map is not uniquely defined  by condition (\ref{SW}). The
ambiguities that arise \cite{AsKi} will be discussed shortly.

The condition (\ref{SW}) yields a simultaneous equation for $A_i$ and
$\lam$.  For the constant $\theta$ case, explicit solutions of the
Seiberg-Witten map have been found by various authors up to second
order in $\theta$~\cite{GoHa, Wess}. The solutions were found by
writing the map as a linear combination of all possible terms allowed
by index structure and dimensional constraints and then determining the
coefficients by plugging this expression into the SW equation. The
method we will describe in the rest of the paper provides a more 
systematic procedure for solving the SW map.
For the special case of a U(1) gauge
group, an exact solution in terms of the Kontsevich formality map is given
in \cite{JuSchWe}, while \cite{Liu, LiMi, Mukhi, LiMi} present an inverse of
the SW map to all orders in $\theta$.

An alternative characterization of the Seiberg-Witten map can be
obtained following \cite{Wess}. In the commutative gauge theory, one may
consider a field $\psi$ in the fundamental representation of the gauge
group. If we assume that the SW map can be extended to include such
fields, then there will be a field $\Psi$ in the noncommutative theory
with the following functional dependence

\beq
\label{map2}
\Psi=\Psi(\psi, \partial \psi, \cdots, a, \partial a, \cdots),
\eeq
and with the corresponding infinitesimal gauge transformation
\beq
\label{psigauge}
\dela \psi= i \al \psi
\eeq
\beq
\label{eqforpsi}
\dell \Psi= i\lam \star \Psi.
\eeq
An alternative to the SW condition~(\ref{SW}) can now be given by

\beq
\label{SW2}
\Psi +\dell\Psi=\Psi(\psi +\dela \psi,\cdots, a_j +\dela a_j, \cdots).
\eeq
More compactly, one writes
\beq
\della \Psi(\psi,a_j,\cdots)=\dela \Psi(\psi,a_j,\cdots).
\eeq
The dependence of $\lam$ on $\al$ is shown explicitly on the left hand side,
and on the right hand side $\dela$ acts
as a derivation on the function $\Psi$, with an action on the variables
$\psi$ and $a_i$ given by (\ref{psigauge}) and (\ref{agauge}) respectively.
Next, one considers the commutator of two infinitesimal gauge transformations
\beq
\label{commutator}
\left [ \della, \dellb \right] \Psi=\left [\dela,\delb \right ]\Psi.
\eeq
Since $ [ \dela, \delb ]= \delta _ {-i[ \al,\beta]}$,
the right hand side of (\ref{commutator}) can be rewritten as
\[
\delta _ {-i[ \al,\beta]}\Psi=\delta_{\lam_{-i[
    \al,\beta]}}\Psi
\\
= i\lam_{-i[ \al,\beta]} \star \Psi=\lam_{[
  \al,\beta]} \star \Psi.
\]
The last equality follows from the fact that $\lam$ 
is linear in the ordinary gauge parameter, which is infinitesimal.
As for the left hand side,
\[
\left [\della, \dellb \right]\Psi=\della \left ( i \lamb \star \Psi \right ) -
\dellb \left ( i \lama \star \Psi \right )
\]
\[=i\left (\dela \lamb -\delb 
\lama \right ) \star \Psi + \left [ \lama \starcom \lamb \right] \star \Psi.
\]
Equating the two expressions and dropping $\Psi$ yields
\beq
\label{eqforl}
\left (\dela \lamb -\delb \lama \right ) -i \left [ \lama \starcom \lamb
\right ]+ i\lam_{[ \al,\beta]}=0~.
\eeq
An advantage of this formulation is that (\ref{eqforl}) is an equation in  
$\lam$ only, whereas (\ref{SW}) must be solved simultaneously
in $\lam$ and $A_i$.
If (\ref{eqforl}) is solved, (\ref{eqfora}) then yields an equation for 
$A_i$ and (\ref{eqforpsi}) for $\Psi$.



\section{The Ghost Field and the Coboundary Operator}
\setcounter{equation}{0}
It is advantageous to rewrite equations (\ref{eqfora}),
(\ref{eqforpsi})
and (\ref{eqforl})
in terms of a ghost field in order to make explicit an underlying
cohomological structure. 
Specifically, we replace the gauge parameter $\al$ with a ghost $v$,
which is an enveloping algebra valued, Grassmannian field\footnote{
In the $U(1)$ case the introduction of a ghost has been considered in
\cite{Oku}.}.  We define a
ghost number by assigning ghost number one to $v$ and zero to $a_i$
and $\psi$. The ghost number introduces a $Z_2$ grading, with even quantities
commuting and odd quantities anticommuting.
In our formalism, the gauge
transformations (\ref{agauge}) and (\ref{psigauge}) are
replaced by the following BRST transformations:
\beq
\label{BRST}
 \begin{array}{ll}
\delv v=iv^2\\
\delv a_i=\partial_i v -i\left [ a_i, v\right ] \\
\delv \psi= iv\psi~.
\end{array}
\eeq
In the $U(1)$ case the introduction of a ghost has been considered in 
\cite{Oku}.
We also take $\delv$ to commute with the partial derivatives,
\beq
\label{ddcommute}
[\delv,\partial_i ] = 0~.
\eeq
The operator $\delv$ has ghost number one and obeys a graded Leibniz rule
\beq
\label{antiLeibniz}
\delv (f_1 f_2)=(\delv f_1) f_2 +(-1)^{deg(f_1)} f_1 (\delv f_2)~,
\eeq
where $deg(f)$ gives the ghost number of the expression $f$.
One can readily check that $\delv$ is nilpotent on the fields $a_i$,
$\psi$ 
and $v$ and therefore, as a consequence of
(\ref{antiLeibniz}), we have
\beq
\label{nilpotent}
\delv ^2=0~.
\eeq

Following the procedure outlined in the previous section, we
characterize the SW map as follows. We introduce a matter field
$\Psi(\psi,\partial \psi, \cdots, a,\partial a, \cdots)$ and an odd
gauge parameter $\lam (v,\partial v, \cdots,  a, \partial a,\cdots)$
corresponding to $\psi$ and $v$ in the commutative theory. 
$\lam$ is linear in the infinitesimal parameter $v$ and hence has
ghost number one. As before, we require that the SW map respect gauge
invariance.
\beq
\label{neweqforpsi}
\dell \Psi  \equiv i \lam \star \Psi =\delv   \Psi. 
\eeq
The consistency condition (\ref {commutator}) now takes the form
\beq
\dell  ^2 \Psi =\delv ^2 \Psi =0~,
\eeq
and again it yields an equation in $\lam$ only.
\[
0=\dell ^2 \Psi =\dell (i \lam \star \Psi)=i \delv \lam \star \Psi +\lam
\star \lam \star \Psi
\]
Since $\Psi$ is arbitrary we obtain
\beq
\label{neweqforl}
\delv \lam = i \lam \star \lam .
\eeq
Once the solution of (\ref{neweqforl}) is known, one can solve the
following equations for~$\Psi$ and the gauge field.
\beq
\label{eqforpsiagain}
\delv \Psi= i \lam \star \Psi~,~~~~~
\delv A_i= \partial_i \lam -i \left [ A_i \starcom \lam \right ]~.
\eeq

It is natural to expand $\lam$ and $A_i$ as power series in the
deformation parameter $\theta$. We indicate the order in $\theta$
by an upper index in parentheses
\beq
\label{expansion}
\begin{array}{ll}
\lam = \sum_{n=0}^{\infty} \lam^{(n)} = v + \sum_{n=1}^{\infty} \lam^{(n)}
\\
A_i= \sum_{n=0}^{\infty} A_i^{(n)} = a_i+ \sum_{n=1}^{\infty} A_i^{(n)}~.
\end{array}
\eeq
Note that the zeroth order terms are determined by requiring that the
SW map reduce to the identity as $\theta$ goes to zero.  
Using this expansion we can rewrite equations~(\ref{neweqforl})
and (\ref{eqforpsiagain}) as

\beq
\label{Zuminobyorders}
\begin{array}{ll}
\delv \lam^{(n)} - i \{ v, \lam^{(n)} \} = M^{(n)} \\ 
\delv A_i^{(n)} - i [ v, A_i^{(n)} ] = U_i^{(n)} ~,
\end{array}
\eeq
where, in the first equation, $M^{(n)}$ collects all terms of order
$n$ which do not contain $ \lam^{(n)}$,and similarly $U_i^{(n)}$
collects 
terms not involving $A_i^{(n)}$.
We refer to the left hand side of each equation as its
homogeneous part, and to $M$ and  $U_i$ as the inhomogeneous
terms of ~(\ref{Zuminobyorders}).  Note that $M^{(n)}$ contains
explicit factors of $\theta$, originating from the expansion of the
Weyl-Moyal product (\ref{WMproduct}). An expression for the generic
$M^{(n)}$ is given in the Appendix. If the SW map for $\lam$ is known
up to order $(n-1)$, then $M^{(n)}$ can be calculated explicitly as a
function of $v$ and $a_i$.  On the other hand, $U^{(n)}$ depends on
both $\lam$ and $A_i$, the former up to order $n$ and the latter up to
order $(n-1)$.  Still, one can calculate it iteratively as a function
of $v$ and $a_i$.

The structure of the homogeneous portions 
suggests the introduction of a new operator $\Delta$.
\beq
\label{Ddef}
\Delta=\left \{\begin{array}{ll}
\delv -i \{v, \cdot\} & \textrm{on odd quantities} \\
\delv -i [v,\cdot ] &  \textrm{on even quantities}\\
\end{array}
\right.
\eeq
In particular, $\Delta$ acts on $v$ and $a_i$ as follows.
\beq
\Delta v=-iv^2~,~~~~
\Delta a_i=\partial_i v~.
\eeq
As a consequence of its definition, $\Delta$ is an anti-derivation
with ghost-number one. It follows a graded Leibniz rule identical to
the one for $\delv$ (\ref{antiLeibniz}). Another consequence of the definition
(\ref{Ddef}) is that $\D$ is nilpotent

\beq
\label{Dnilpotent}
\Delta^2=0~.
\eeq

The action of $\Delta$ on expressions involving $a_i$ 
and its derivatives can also be characterized in geometric terms.
Specifically, $\Delta$ differs from $\delv$ in that it removes the covariant
part of the gauge transformation. 
Therefore, $\Delta$ acting on any covariant expression will give zero. 
For instance, if one constructs the field-strength,
$F_{ij} \equiv \partial_i a_j - \partial_j a_i - i [a_i, a_j]$, 
one finds by explicit calculation
\beq
\label{ddf}
\Delta F_{ij}=0.
\eeq
It can also be checked that the covariant derivative,
\mbox{$D_i=\partial_i -i[a_i, \cdot]$} commutes with $\Delta$,
\beq
[\Delta,D_i]=0.
\label{covdelta}
\eeq

In terms of $\Delta$ the equations (\ref{Zuminobyorders}) take the form
\beq
\label{cohomology}
\begin{array}{ll}
\Delta \lam^{(n)}= M^{(n)} \\ 
\Delta A_i^{(n)}= U_i^{(n)}~.
\end{array}
\eeq
In the next section we will provide a method to solve
these equations. 
Also note that since $\Delta^2=0$, it must be true that
\beq
\label{closed}
\begin{array}{ll}
\Delta M^{(n)}=0 \\ 
\Delta U_i^{(n)}=0~. 
\end{array}
\eeq
Indeed one should verify that (\ref{closed}) holds order by order. If
(\ref{closed}) did not hold, this would signal an inconsistency in
the SW map.

\section{The Homotopy Operator}

For simplicity, 
we begin by considering in detail the SW Map for the case of the gauge
parameter $\Lambda$. Much of what we say actually applies to the other
cases as well with minor modifications.


In the previous section, we have seen that order by
order in an expansion in $\theta$, the SW map has the form:
\beq
\label{lameq}
\Delta \Lambda^{(n)} = M^{(n)},
\eeq
where $M^{(n)}$ depends only on $\Lambda^{(i)}$ with $i<n$. Clearly, if one
could invert $\D$ somehow, we could solve for $\L$. But $\D$ is
obviously not invertible, as $\D^2=0$. In particular, the solutions of
(\ref{lameq}) are not unique, since if $\L$ is a solution so is
$\L + \D S$ for any $S$ of ghost number zero\footnote{These are precisely the ambiguities in the SW
  map that were first discussed in \cite{AsKi}, where our operator
  $\D$ was called $\hat{\delta\p}$.}. That is, $\D$ acts like a
coboundary operator in a cohomology theory, and the solutions that we
are looking for are actually cohomology classes of solutions, unique
only up to the addition of $\D$-exact terms. 
The formal existence of the SW Map is then equivalent to the statement that
the cycle $M\on$ is actually $\D$-exact for all $n$. 
Since we know
that $\D^2=0$, this fact would follow as a corollary of the stronger
statement that there is no non-trivial $\D$-cohomology in ghost number
two. In other words, there are no $\D$-closed, order $n$ polynomials
with ghost number two which are not also $\D$-exact. To prove this
stronger claim, we could proceed as follows. Suppose that we could
construct an operator $K$ such that 
\beq
K\D + \D K = 1.
\eeq
Clearly, $K$ must reduce ghost number by one, and therefore must be
odd. Consider its action on a cycle $M$, (so $\D M = 0$)
\beq
(K \D + \D K)M = \D K M = M.
\eeq
Therefore, $M = \D \L$, with $\L = K M$, which not only shows that $M$
is exact, but also computes explicitly a solution to the SW map. We
note that this method of solution is nearly identical to the method
used by Stora and Zumino~\cite{Zumino} to solve the Wess-Zumino consistency
conditions for non-Abelian anomalies. In fact, it was the parallels
between these problems that motivated our current approach. 

We now proceed to construct $K$.
First we notice that $M^{(n)}$ depends on $v$ only through 
its derivative $\partial_i v$, as one can see by looking at
the explicit expressions in the Appendix. 
The same is true for 
$U_i^{(n)}$ since it depends on $v$ only
through $\lam$.
It  is convenient to define
\be
\label{defb}
b_i=\partial_i v~,
\ee
so that $M$ and  $U_i$ can all be expressed as functions of $a_i$,
$b_i$ and their derivatives only. Furthermore,
we rewrite $M\on$ solely in terms of covariant
derivatives, rather than ordinary ones. After these replacements,
we may consider $M\on$ an
element of the algebra generated by $a_i$, $b_i$, and $D_i$. As
explained
in the next section this algebra is not free, but for the
moment we ignore this issue.
The action of the operator $\Delta$ takes on a particularly
simple form in terms of these variables: 
\beq
\label{Daction}
\D a_i=b_i~,~~~
\D b_i=0 ~,~~~
\left[ \D,D_i \right]=0.
\eeq 
Thus, it is natural to define $K$ on these variables. A natural guess is

\beq
\label{Kaction1}
K a_i=0~,~~~
K b_i=a_i.
\eeq 

Since $K$ inverts an operator which acts like a graded
derivation, it cannot itself obey the
Leibniz rule.  We can instead proceed by defining an
infinitesimal form of the operator $K$, which does. In particular, to
define $K$, we first define two operators $\k$ and $\d$ such that 
\beq
\label{ed}
\D \k + \k \D= \d ,
\eeq
and then an operator $T$ (a kind of integration operator) such that
\beq
\label{Taction1}
T \d M = M~,~~~ 
T(\k M) = KM.
\eeq 
The operator $\d$ is some infinitesimal variation of $a_i$ and $b_i$,
which can be integrated to the identity. It is also defined
to commute with the covariant derivative
\beq
\left[\d,D_i\right]=0.
\eeq 
The action of $\k$ is defined by
\beq
\label{laction}
\k a_i=0~,~~~
\k b_i=\d a_i~,~~~
\left[\k,D_i\right]=0~,~~~
\left[\k,\d \right]=0~,~~~
\k^2=0.
\eeq 
Finally, the integration operator $T$ acting on any expression is implemented
via the following procedure:
\begin{enumerate}
\item Choose the fields to be linearly dependent
on $t$ and $\d$ to be the infinitesimal variation with respect to t:
\beq
\label{daction}
\begin{array}{ll}
\d a_i \rightarrow a_i dt \\
\d b_i \rightarrow b_i dt \\
a_i \rightarrow t a_i \\
b_i \rightarrow t b_i \\
\end{array}
\eeq 
That is, we transform any expression,
\beq
N(a_i,b_i,\d a_i,\d b_i,D_i) \rightarrow N(ta_i,tb_i,a_i dt,b_i dt,
D_i).
\eeq
\item Integrate from $t=0$ to $t=1$. Thus,
\beq\label{Taction2}
T N(a_i,b_i,\d a_i,\d b_i,D_i) = \int_{0}^{1} N(ta_i,tb_i,a_i
dt,b_i dt, D_i).
\eeq
\end{enumerate}
Notice that this prescription requires that we rewrite any expression
involving ordinary derivatives in terms of covariant derivatives and
gauge fields only. 
We now show by induction that these definitions do in fact yield a
homotopy operator $K$. 
It is easy to see that $\D \k + \k \D = \d$ holds when acting on
$a_i$ or $b_i$ alone. Suppose then that the equation holds when
acting on two monomials $f$ and $g$ of order less than or equal to $r$ 
in $a_i$ and $b_i$. Then it
follows that
\beq
(\D \k + \k \D)(fg) = ((\D \k + \k \D)f)g + f(\D \k + \k \D)g~,
\eeq
where all the cross terms have canceled out. By the induction 
hypothesis this
expression is equal to $(\d f)g + f\d g$, which is just
$\d (fg)$. Thus $\D \k + \k \D = \d$ holds on any monomial of degree
greater than zero. Since this operator is distributive, (\ref{ed}) holds
for any element of the algebra.


\section{Constraints}

We have
so far only considered the free algebra, generated by
$a_i$\,, $b_i$ and $D_i$, where the construction
of $K$ was relatively simple. To show that our algebra is not free
consider the following.
\beq
\begin{array}{ll}
\D F_{ij} & = \D\left( D_i a_j - D_j a_i + i[a_i, a_j]\right). \\
& =  D_i b_j - D_j b_i + i[b_i, a_j] + i[a_i, b_j]~.
\end{array}
\eeq
As an element of the free algebra, the left hand side is not zero,
but according to~(\ref{ddf}) it should be.
The problem becomes more serious when one rewrites $M\on$ in terms
of the elements of the free algebra. Beyond first order,
one finds the
$\D M$ is no longer zero in general, but vanishes only by using
the following constraints
\beq
\label{con20}
\left[ F_{ij},
  \cdot\, \right] - i [D_i, D_j](\cdot) = 0~,~~~
\D F_{ij} = 0~.
\eeq
If $\D M\on$ is not zero identically, $K$ no longer inverts $\D$ when
acting on  $M\on$, and we no longer have a method for
  solving~(\ref{cohomology}) for $\L \on$.
The origin of the constraints can be traced to 
the fact that partial derivatives commute
\beq\label{const10}
\del_i \del_j - \del_j \del_i=0~,~~~ 
\del_i b_j - \del_j b_i=0, 
\eeq 
since $b_i = \del_iv$. This is no longer manifest in our algebra. 
In fact, written in terms of covariant derivatives,
(\ref{const10}) becomes (\ref{con20}). There seems to be no way to
eliminate these
constraints since $K$ is not defined on $v$, but only on  $b_i = \del_iv$. 
One might expect that at higher orders one would have to use
  additional constraints to verify  that $\D M\on$ vanishes, but this 
  is not the case.
For example, when one
rewrites
\beq\label{con3}
\del_i\del_k b_j - \del_j\del_k b_i=0 
\eeq 
in terms of covariant derivatives, the resulting expression is not
an independent constraint, but can be written in terms of
the two fundamental ones (\ref{con20}).

The reason why $\D M\on$ is not zero in general is because the existence
of the constraints allows us to write $M\on$ in terms of the
algebra elements in many different ways. Our goal will then be
to define a procedure for writing $M\on$ in terms of algebra elements
so that $\D M\on = 0$, identically. We will describe two procedures.

The first is the method we will use in the next section
to calculate some low
order terms of the SW map. We begin by obtaining an expression for
$M\on$ in terms of the algebra elements. Generically, $\D M\on $ will be
proportional to the constraints. At low
orders, once $\D M\on $ is calculated, it is easy to guess an expression
$m \on $, which is proportional to the constraints, such that the
combination $M\on + m\on $ is annihilated by $\D$. Acting $K$ on this
new combination then gives the solution $\L \on$. We believe this guessing method
can be formalized, but at higher orders we believe that the second
procedure which we will now describe is simpler.

First we introduce
a new element of the algebra, $f_{ij}$, which is annihilated by all the
operators defined in previous sections.
\beq
\D f_{ij} = \d f_{ij} = \k f_{ij} = 0~.
\eeq
We also introduce a new constraint
\beq
\label{newcon}
f_{ij} - F_{ij} = 0~,
\eeq
where $F_{ij}$ is considered a function of $D_i$ and $a_i$.
We want to show that using this enlarged algebra and the constraints
we can rewrite $M\on$ so that it has the following dependence.
\beq
\label{totsym}
M\on = M\on( a, b, (D^ka)_s, (D^lb)_s, D^hf)~,
\eeq
where the subscript $s$ indicates that all the indices within the
parentheses should be totally symmetrized. It would then follow that
$\D M$ has the same functional dependence. Since it is impossible that
$\D M$ contains any term antisymmetric in the indices of $Da$ or $Db$,
the constraints (\ref{newcon}) and (\ref{con20}) cannot be generated.
However, we may find that $\D M$ is proportional to the following
constraints. 
\beq
\label{nnewcon}
\left[ f_{ij},
  \cdot\, \right] - i [D_i, D_j](\cdot) = 0~,~~D_if_{jk} +D_jf_{ki} +
  D_kf_{ij} =0~.
\eeq
Since these constraints commute with the action of both $K$ and
$\D$, if we add to $M$ a term proportional to (\ref{nnewcon}), our
result for $\L = KM$ is unchanged. To show that we can actually
write $M$ in the form suggested above, we begin with an expression
for $M$ as found by expanding the star product.
\beq
M\on = M\on( a , (\del^k)_s a , (\del^l)_s v )~,
\eeq
where we choose to explicitly write the derivatives in symmetric
form. By replacing $\del (\cdot )\rightarrow D (\cdot )+ i[a,\cdot]$,
and $\del v \rightarrow b$ the expression takes the form
\beq 
M\on = M\on( a, b, (D^k)_sa, (D^lb)_s)~.
\eeq
The difference $(D^ka)_s - D^ka$ contains terms that are
proportional to the antisymmetric parts of $DD$ or $Da$. But using
the constraints we can make the following substitutions
\beq
\label{substitute}
[D_i,D_j](\cdot) \rightarrow -i[f_{ij}, \cdot\,]~,~~~D_i a_j - D_j a_i \rightarrow 
f_{ij} -i[a_i,a_j]~.
\eeq
This must be done recursively since the commutator term involving
$a$'s above
may again be acted on by $D$'s.
But at each step, the number of possible $D$'s acting on $a$ is
reduced by one. 
After carrying out this procedure $M$ will have
the form (\ref{totsym}).


\section{Some Calculations} 

In this final section, we use the formalism we have developed to 
compute some low order terms of the SW map. We focus mainly on solving
for the gauge parameter $\L$. 


At the zeroth order, if we expand 
$
\delv \lam = i \lam \star \lam$
we find
\ba
\delv v&=i v^2 
\ea
which is just the BRST transformation of  $v$ (\ref{BRST}).
At first order, we have
\beq
\Delta \Lambda^{(1)}=-\frac{1}{2} \t^{ij} b_i b_j
\label{m1}
\eeq
while at the second order we obtain
\beq
\Delta \Lambda^{(2)}=-\frac{i}{8} \t^{ij} \t^{kl} \partial_i b_k \partial_j b_l
-\frac{1}{2} \t^{ij} [b_i,\partial_j \Lambda^{(1)}] + i \Lambda^{(1)} \Lambda^{(1)}~.
\label{m2}
\eeq
A solution of (\ref{m1}) has been found in \cite{SW} and is given by
\beq
\Lambda^{(1)}=\frac{1}{4} \t^{ij} \{b_i,a_j\}~.
\eeq
We can reproduce this solution immediately by applying $K$ to the expression
$M^{(1)}=-\frac{1}{2} \t^{ij} b_i b_j$. There
are no problems at this level, since there are not enough derivatives
for the constraints to show up.
As explained in the previous section we proceed in two steps. We first
apply $\ell$
\beq
\ell (M^{(1)})=-\frac{1}{2} \t^{ij} (\delta a_i b_j - b_i \delta a_j)~,
\eeq
then $T$ to find
\ba
K (M^{(1)})=\frac{1}{2}\theta^{ij} (b_i a_j +a_j b_i) \int_0^1 dt t
=\frac{1}{4}\theta^{ij} \{b_i, a_j \}~.
\ea
The ambiguity in the first order solution as determined in 
\cite{AsKi} is proportional to
\beq
\tilde \Lambda^{(1)}=-2 i\t^{ij} [b_i,a_j]~.
\eeq
According to the previous discussion the ambiguity amounts to an exact
cocycle, hence is of the form:
\beq
\tilde \Lambda^{(1)}=\Delta S^{(1)}~,
\eeq
where $S^{(1)}$ can be computed to be
\beq
S^{(1)}=K \tilde \Lambda^{(1)}=-i \t^{ij}[a_i,a_j]~.
\eeq
Solutions at the second order have been found by various authors.
In \cite{Wess} the following solution is presented,
\footnote{The expression for $\L^{(2)}$ found in~\cite{Wess} appears to
  contain a misprint. The correct formula~(\ref{l2w})
  was in
  fact given to us personally by J. Wess.}
\ba
\label{l2w}
\Lambda^{(2)}&=&\frac{1}{32} \t^{ij} \t^{kl} 
\Big(-\big\{b_i,\{a_k,i [a_j,a_l]+4 
\partial_l a_j\}\big\} -i \big\{a_j,\{ a_l,[b_i, a_k]\}
\big\} \label{Munich} \nn \\
&&+2 \big[ [b_i,a_k]+i \partial_i b_k ,\partial_j a_l \big] +2i
\big[ [a_j,a_l],[b_i,a_k]\big]\Big)~,
\ea
while in \cite{GoHa} the following solution is found,
\ba
{\Lambda'}^{(2)}&=&\frac{1}{32} \t^{ij} \t^{kl} 
\Big(-\big\{b_i,\{a_k,i [a_j,a_l]+4 
\partial_l a_j\}\big\} -i \big\{a_j,\{ a_l,[b_i, a_k]\}
\big\} \nn \\
&&+2 \big[ [b_k,a_i]+i \partial_i b_k ,\partial_j a_l \big]\Big)~. 
\ea
According to our previous observation the difference between these two
expressions must be of the form $\Delta S^{(2)}$.
In fact, we find
\beq
\Lambda^{(2)}-{\Lambda'}^{(2)}=\frac{1}{16}\t^{ij} \t^{kl} 
\Big(\big[\partial_i a_k,[b_l,a_j]+[a_l,b_j] \big]
-i \big[[a_k,b_i],[a_l,a_j]\big]\Big)=\Delta S^{(2)},
\eeq
with $S^{(2)}$ given by
\beq
S^{(2)}=\frac{1}{16}\t^{ij} \t^{kl}\big[\partial_i a_k,[a_l,a_j] \big]~.
\label{SWJ}
\eeq
This expression for $S^{(2)}$ can be obtained in the following way.
According to the prescriptions in the previous section, before we can
apply $K$ to $\Lambda^{(2)}-{\Lambda'}^{(2)}$, we need that
$\Delta(\Lambda^{(2)}-{\Lambda'}^{(2)})$ vanishes algebraically, without 
using the relations (\ref{con20}).
We observe that
\beq
\Delta (\Lambda^{(2)}-{\Lambda'}^{(2)})=-\frac{1}{16} \t^{ij} \t^{kl} 
\{\Delta F_{ik},a_j b_l+b_j a_l\}=\Delta (\frac{1}{16} [\Delta F_{ik},a_j a_l])
\eeq
which vanishes only by means of the constraint 
$\Delta F_{ij}=0$.
Therefore we add to $\Lambda^{(2)}-{\Lambda'}^{(2)}$ a term
$-\frac{1}{16} \t^{ij} \t^{kl} [\Delta F_{ik},a_j a_l]$ 
and only at this point we apply $K$, which yields (\ref{SWJ}).

To ensure that $\Delta (\Lambda^{(2)}-{\Lambda'}^{(2)})$ vanishes, 
we could have also 
used our other prescription to symmetrize 
$\Lambda^{(2)}-{\Lambda'}^{(2)}$ with respect to all derivatives and then
use the substitution~(\ref{substitute})
\beq
 f_{ij} - D_i a_j - D_j a_i + i[a_i, a_j]
  = 0,
\eeq
to replace $F$ with $f$.

\ba
\Lambda^{(2)}-{\Lambda'}^{(2)}&=&\frac{1}{16}\t^{ij} \t^{kl} 
\Big(\big[F_{ik}+i[a_i,a_k],[b_l,a_j] \big]
+i \big[[a_k,b_i],[a_j,a_l]\big]\Big) \nn \\
&=&\frac{1}{16}\t^{ij} \t^{kl} [f_{ik},[b_l,a_j]]~.
\ea
By applying $K$ we immediately get
\beq
S^{(2)}=K (\Lambda^{(2)}-{\Lambda'}^{(2)})=\frac{1}{32}\t^{ij} \t^{kl} [f_{ik},[a_l,a_j]]
\eeq
By substituting back the expression for $f_{ik}$ and noting that
\beq
\theta^{ij}\theta^{kl}\big[[a_k,a_i],[a_j,a_l]\big]=0
\eeq
we again recover (\ref{SWJ}).

By following the same procedure we can compute directly a solution of
(\ref{m2}) at the second order.
\ba
{\Lambda''}^{(2)}&=&-\frac{1}{2} \t^{ij} \big\{a_i,\frac{1}{3} D_j \Lambda^{(1)}
+\frac{i}{4} [a_j,\Lambda^{(1)}] \big\}+\t^{ij} \t^{kl} \Big(-\frac{i}{16}
[D_i a_k,D_j b_l] \no
&&+ \big[ [a_i,a_k],  \frac{1}{24} D_j b_l 
+\frac{i}{32} [a_j,b_l] \big] +\frac{1}{24} \big[D_ia_k,[a_j,b_l]\big]\no
&&+\frac{1}{8} \Big(a_i (\frac{1}{3} 
D_j a_k-\frac{1}{3} D_k a_j +\frac{i}{2} [a_j,a_k] \big) b_l \label{our}\\
&&-b_i (\frac{1}{3} D_j a_k-\frac{1}{3} D_k a_j 
+\frac{i}{2} [a_j,a_k]) a_l \no
&&+\big\{ \frac{1}{6} (D_i a_k-D_k a_i)+\frac{i}{4}[a_i,a_k],
\{a_l,b_j\} \big\}\Big) \Big)~. \nn
\ea
To obtain this result we first observe that
\beq
\Delta M^{(2)}=\frac{1}{8} \t^{ij} \t^{kl}
\Big(-2 b_i \Delta F_{jk} b_l+  b_i b_k
\Delta F_{lj} +\Delta F_{ik} b_l b_j \Big)~.
\eeq
Again, as $\Delta M^{(2)}$ vanishes only due to the constraint, we add
\beq
{m}^{(2)}=\frac{1}{16} \t^{ij} \t^{kl} \Big(2 a_i \Delta F_{jk} b_l
+2 b_i \Delta F_{jk} a_l - (a_i b_k-b_i a_k) \Delta F_{lj} -
\Delta F_{ik} (b_l a_j -a_l b_j) \Big)
\eeq
in such a way as to obtain $\Delta (M^{(2)}+{m}^{(2)})=0$.
Notice that there is an ambiguity in the choice of ${m}^{(2)}$, but we have
chosen the particular  ${m}^{(2)}$  which respects the reality structure,
i.e. which provides us with a real $\Lambda''_2$.
Moreover, observe that
\beq
\k \Lambda^{(1)}=0, \quad \k \partial_i \Lambda^{(1)}=0~.
\eeq
This is a consequence of the fact that
\beq
\k K=0
\eeq
and
\beq
\Lambda^{(1)}=K M^{(1)}~.
\eeq
As expected, the difference between our solution ${\Lambda''}^{(2)}$ 
(\ref{our}) and the solution $\Lambda^{(2)}$
(\ref{Munich}) is again of the form ${\Delta S\p}^{(2)}$ (up to a term which vanishes 
by the constraint).
\ba
{\Lambda''}^{(2)}-\Lambda^{(2)}&=&\theta^{ij}\theta^{kl}\Big[\Delta \Big( \frac{1}{24} 
(\big[a_j,[D_i a_k,a_l] \big]+2 (D_i a_k a_j a_l+
a_l a_j D_i a_k \Big) \no
&&+\frac{1}{16} [a_i a_k,\Delta F_{jl}]\Big]~.
\ea
A similar technique can be followed for the potential $A_i$.
Moreover, if $\Lambda \on$ is changed by an amount $\Delta S^{(n)}$
\beq
\Lambda^{(n)} \rightarrow \Lambda^{(n)}+\Delta S^{(n)}
\eeq
then the corresponding change in the potential is 
\beq
A^{(n)}_i \rightarrow A^{(n)}_i+D_i S^{(n)}~.
\label{ambA}
\eeq
This follows from the fact that the equation of order $n$ for the gauge field is always of the
form
\beq
\Delta A^{(n)}_i=D_i \Lambda^{(n)} +\cdots
\label{eqA}
\eeq
Notice that
(\ref{ambA}) is a consequence of the fact that
the coboundary operator $\Delta$ commutes with the covariant derivative $D_i$.

\section*{Appendix}

In this appendix we give some useful expressions arising from the
expansion of the Weyl-Moyal product. 

First, for simplicity, we will define 
\ba
\del_{I_n} = \del_{i_1} \cdots \del_{i_n} \\
\t^{I_n J_n} = \t^{i_1 j_1} \cdots \t^{i_n j_n}~.
\ea
We will expand out $*$-products using Moyal's formula:
\ba
f(x) * g(x)
&= &e^{\frac{i}{2} \theta^{ij} \del_y \del_z} f(y)g(z) |_{y=z=x} \no
&=&\sum_{n=0}^{\infty} \frac{1}{n!} \left(\frac{i}{2} \right)^n
\t^{i_1 j_1} \cdots \t^{i_n j_n} \del{i_1} \cdots \del_{i_n} f(x) 
\del_{j_1} \cdots \del_{j_n} g(x) \no
&=&\sum_{n=0}^{\infty} \frac{1}{n!} \left(\frac{i}{2} \right)^n \t^{I_n J_n}
\del_{I_n} f(x) \del_{J_n} g(x)~.
\ea

Inserting this expansion into (\ref{neweqforl}) and requiring that the
equation is satisfied order by order in $\theta$, we find the
following expression
\ba
\lefteqn{\delv \Lambda^{(n)} - i \left\{ \Lambda^{(n)}, v \right\}
= \Delta \Lambda^{(n)}}\\
&& =\sum_{p=1}^{n-1} \left( \frac{i}{(n-p)!} \left(\frac{i}{2} \right)^{n-p}
\t^{I_{n-p} J_{n-p}} \left\{ \del_{I_{n-p}}
    \Lambda_p , \del_{J_{n-p}} v \right\} \right. \no
&& \left. + \frac{i}{(p-1)!} 
\left(\frac{i}{2} \right)^{p-1} \t^{I_{p-1} J_{p-1}} 
\sum_{q=1}^{n-p} \del_{I_{p-1}} \Lambda_q
\del_{J_{p-1}} \Lambda_{n-q-p+1} \right) \no
&&+\frac{i}{n!} \left(\frac{i}{2} \right)^n \t^{I_n J_n} 
\left( \del_{I_n} v \right) \left( \del_{J_n} v \right)
\nn
\ea
Up to the second order this equation reads
\ba
{0^{th} :} &&\delv v=i v^2\\
{1^{st} :}& &\Delta \Lambda^{(1)}=-\frac{1}{2} \t^{ij} b_i b_j \\
{2^{nd }:}&&\Delta \Lambda^{(2)}=-\frac{i}{8} \t^{ij} \t^{kl} 
\partial_i b_k \partial_j b_l
-\frac{1}{2} \t^{ij} [b_i,\partial_j \Lambda^{(1)}] + i  \Lambda^{(1)}
\Lambda^{(1)}~.
\ea
Analogously the equation (\ref{eqforpsiagain}) for the gauge potential $A_i$
\beq
\delv A_i= \partial_i \lam -i \left [ A_i \starcom \lam \right ]
\eeq
reads
\ba
{0^{th} :} &&\Delta A_i^{(0)}=b_i\\
{1^{st} :}&& \Delta A_i^{(1)}=D_i \Lambda^{(1)}-\frac{1}{2} \t^{ij}
\{b_k,\del_l a_i \}\\
{2^{nd }:}&& \Delta A_i^{(2)}=D_i \Lambda^{(2)} +i [\Lambda^{(1)},
A_i^{(1)}]-\frac{1}{2} \t^{kl} \{b_k,\del_l A_i^{(1)} \} \no
&&-\frac{1}{2} \t^{kl} \{ \del_k \Lambda^{(1)},\del_l a_i \}-\frac{i}{8}
 \t^{kl} \t^{mn} [\del_k b_m,\del_l \del_n a_i]~.  
\ea

\setcounter{section}{1}
\def\thesection{\Alph{section}}
\setcounter{equation}{0}

\section*{Acknowledgments}

We are very grateful to Julius Wess. A seminar he gave in Berkeley on
the
content of~\cite{Wess}  was the original inspiration for our work.
This work was supported in part by the Director, Office of Science,
Office of High Energy and Nuclear Physics, Division of High Energy Physics of 
the U.S. Department of Energy under Contract DE-AC03-76SF00098 and 
in part by the
National Science Foundation under grant PHY-95-14797. B.L.C. is supported
by the DFG (Deutsche Forschungsgemeinschaft) under grant CE 50/1-1.

\end{document}